\documentclass{appolb}
\usepackage{epsfig}

\def\qsq{q^2}
\def\fp{f_{+}}
\def\dkln{D^{0} \rightarrow K^{-} l^{+} \nu}
\def\dpiln{D^{0} \rightarrow \pi^{-} l^{+} \nu}

\def\dkmn{D^{0} \rightarrow K^{-} \mu^{+} \nu}

\def\dksln{D^{+} \rightarrow \overline{K^{*0}} l^{+} \nu}
\def\dksen{D^{+} \rightarrow \overline{K^{*0}} e^{+} \nu}
\def\dksmn{D^{+} \rightarrow \overline{K^{*0}} \mu^{+} \nu}
\def\dkpimn{D^{+} \rightarrow K^{-}\pi^{+} \mu^{+} \nu}
\def\dsphiln{D_s^{+} \rightarrow \phi l^{+} \nu}
\def\dsphimn{D_s^{+} \rightarrow \phi \mu^{+} \nu}
\def\dkzmn{D^{+} \rightarrow \overline{K^{0}} \mu^{+} \nu}
\def\dkzen{D^{+} \rightarrow \overline{K^{0}} e^{+} \nu}
\def\dkpipi{D^{+} \rightarrow K^{-}\pi^{+} \pi^{+}}

\def\rv{r_{V}}
\def\rtwo{r_{2}}
\def\mpole{\mathrm{M(pole)}}
\def\mpolek{\mathrm{M(pole)}^{D\rightarrow K}}
\def\mpolepi{\mathrm{M(pole)}^{D\rightarrow \pi}}

\def\PRL{Phys.\ Rev.\ Lett.}
\def\PLB{Phys.\ Lett.\ B}
\def\NP{Nucl.\ Phys.\ B}

\begin{document}
\title{Review of Charm Semileptonic Decays and QCD%
 \thanks{Presented at XXXIV International Symposium on Multiparticle Dynamics,
 July 2004, Sonoma County, California, USA}
}
\author{Doris Yangsoo Kim%
 \address{Loomis Lab of Physics, 1110 W Green St., University of Illinois,\\ 
 Urbana, IL., 61801, USA}
}
\maketitle

\begin{abstract}
In this paper, we review recent progress in the field of 
semileptonic decays of charm mesons, including topics on the relative
branching ratio and the form factors. The comparison between 
the experimental form factor measurements and the Lattice QCD calculations is 
emphasized.
\end{abstract}

  
\section{Charm semileptonic decays as tests of Lattice QCD}
The semileptonic decays of charm mesons provide an ideal environment to
refine the QCD physics.  The decay rates are computed from
the first principles using Cabbibo-Kobayashi-Maskawa quark-mixing matrix
elements. The hadronic complications are contained in the form factors,
which are calculable via non-perturbative Lattice QCD, HQET or quark models. 
With the rapid advance in the computer technology, the lattice community
is generating visible improvements in major QCD topics. By comparing the
experimental measurements on the charm semileptonic decays with the lattice
QCD calculations, we can establish a high quality lattice calibration and
reduce systematic errors in the Unitary triangle. 
The QCD techniques validated in the charm decays can be applied to the similar
physics topics in the beauty decays, which will definitely improve
the precision analysis tools to deal with the excellent data sets generated
in the current and future B experiments.

The field of the semileptonic charm physics is quite active. Various new results
have been reported recently and 
several more papers are expected to be published in near future.
In the following sessions, some of the results and the corresponding
theories are summarized. The charge conjugate modes are implicitly
included in all the decay channels mentioned in the paper.

\section{The pseudoscalar channel, $D \rightarrow P l \nu$}

\subsection{The form factors and their parameterization} 
When a charm meson decays into a single pseudoscalar meson, a lepton and 
a neutrino, its decay rate can be described by a simple equation of
$\qsq$ with an easy-to-extract form factor $\fp$ as follows, 

\begin{equation}
\frac{d\Gamma (D \rightarrow P l \nu)}{d\qsq} =
   \frac{G_F^2 |V_{cq}|^2 P_p^3}{24\pi^3}
   ( |\fp(\qsq)|^2 + O(m_l^2) )
\end{equation}

\begin{figure}
\begin{center}\label{lqcd}
\psfig{figure=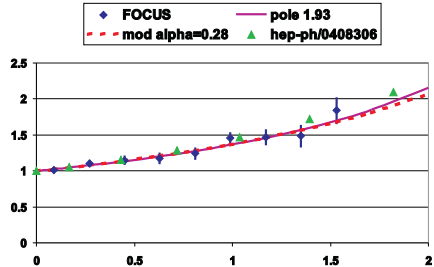,height=1.5in}
\end{center}
\caption{The overlay of the unquenched LQCD calculation of the form factor
$\fp(\qsq)$\cite{lqcd} over the preliminary FOCUS preliminary result. The solid
line represents a pole form fit to the FOCUS data while the dotted line
represents a modified pole form fit. All the data points and the fitted lines
are normalized to have $\fp(0)=1$.}  
\end{figure}

In the Lattice QCD, the $\fp(\qsq)$ distribution is easiest to calculate
when $\qsq = \qsq_{max}$, since the pseudoscalar meson is at rest in the parent
charm meson center of mass frame, which makes the wavelength of the the child
quark larger than the lattice size.  But the decay rate of the charm meson is
smallest at $\qsq = \qsq_{max}$, which suppresses the sensitivity to the $\fp$
measurement. Vice versa, the decay rate of the charm meson is highest
when $\qsq$ is smallest, i.e, where the theoretical calculation is
least certain.  The lattice community is actively working to overcome this
problem and reported a remarkably precise result (See Figure~\ref{lqcd}.)
\cite{lqcd}.

In the past, one used parameterization to describe the $\fp(\qsq)$ distribution,
since the statistics of the experimental data was not large enough. 
There are several parameterization forms in current use:
The more traditional pole form ($\sim 1/(\qsq - \mpole^2)$),
the ISGW1 form ($\sim \exp(\alpha\qsq)$) and its revision ISGW2 form.\footnote{
A recent addition is the modified pole form, which incorporates a second pole
effect.}
The basis of the pole form is as follows: When a charm quark
decays into a lighter quark, the parent and the child quark make
a spin one resonance during the process, which subsequently annihilates
into a $W$ boson. The character of the spin one resonance manifests itself
as the pole mass.

Two decay modes have been studied in the pseudoscalar channel, $\dkln$ and
$\dpiln$. The $\dkln$ decay has the merit of larger branching fraction while
the $\dpiln$ decay has a broader $\qsq$ range, which gives
it a more discerning power among several parameterization models.

\subsection{The new results from 
$D^{0} \rightarrow K^{-} l^{+} \nu / \pi^{-} l^{+} \nu$ decays}

The CLEO experiment reported preliminary results on
the $\dkln$ and the $\dpiln$ decays~\cite{cleo04}, based on the
$7 fp^{-1}$ of data collected with the CLEO III detector. From the pole form
fit to the data, they obtained $\mpolek = 1.89 \pm 0.05 ^{+0.04}_{-0.03} \ GeV$
and $\mpolepi = 1.86 ^{+0.10+0.07}_{-0.06-0.03} \ GeV$. 
The ISGW2 form fit to the $\dkln$ channel was found to be disfavored by 
$4.2 \sigma$. The relative branching ratio between $\dpiln$ and $\dkln$ was
obtained as $0.082 \pm 0.006 \pm 0.005$ and subsequently, 
\begin{equation}
\frac{|\fp^{\pi}(0)|^2 |V_{cd}|^2}
     {|\fp^{K}(0)|^2 |V_{cs}|^2} 
              = 0.038^{+0.006+0.005}_{-0.007-0.003},
\end{equation}
where the first error is the statistical and the second error is systematic.
If the average numbers for the CKM element the Particle Data Group (PDG) are
applied, we get $|\fp^{\pi}(0)| / |\fp^{K}(0)| = 0.86 \pm 0.07 \pm 0.05
\pm 0.01$, which confirms the SU(3) symmetry breaking in the charm meson sector.

At the recent DAFNE04 conference, the BELLE experiment reported a 
preliminary study on the $\dkln$ and $\dpiln$ decays, based on the
$ 152 fp^{-1}$ of data\cite{bellepre}. To obtain a better $q^2$
resolution, they required the candidate events consist of a prompt $D^*$,
a prompt $\overline{D^*}$ and lighter mesons generated from the primary
interaction point.  They concluded that it's possible to get a very
high precision measurement on the $\qsq$ using the BELLE data set,
comparable to the one achievable at a threshold charm factory such
as the CLEO-c. 

\begin{figure}\label{kpoleWA}
\begin{center}
\psfig{figure=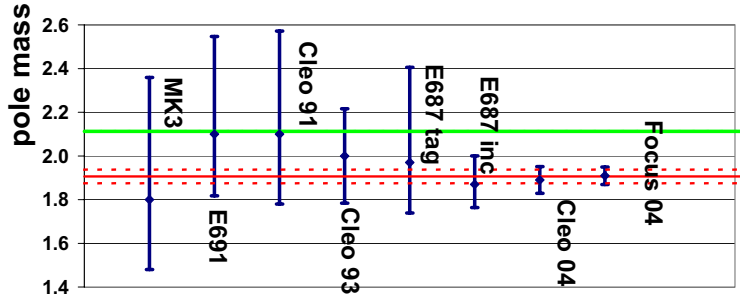,height=2.0in}
\end{center}
\caption{The compilation of the $\mpolek$ including the preliminary CLEO and
FOCUS results. The new world average is calculated as $1.91 \pm 0.03$, shown
as the lower solid line and two dotted lines in the figure. The upper solid
line represents the mass of $D_{s}^{*}$, the pole mass from a naive pole
form model.}
\end{figure}

The FOCUS experiment is also analyzing these decays. They reconstructed
about 12,000 events in the $\dkmn$ decay and obtained a non-parametric
distribution of the $\fp$ form factor (See Figure~\ref{lqcd}.). 

Figure~\ref{kpoleWA} shows the new world average of the $\mpolek$ including
the preliminary results from the CLEO and the FOCUS,  obtained as
$1.91 \pm 0.03$.

\section{The vector channel, $D \rightarrow V l \nu$}
\subsection{Kinematics of the vector decays}
Several decay modes are available for the study of a charm meson decaying
into a vector meson, a lepton and a neutrino. The Cabbibo
allowed $\dksln$ decay is an excellent mode to study, which has
the statistical advantage over others. Since the child $\overline{K^{*0}}$
promptly decays into a $K^-$ and a $\pi^+$, the kinematics
of the $\dksln$ channel becomes 4-body and is described by two invariant
masses and three decay angles.  The decay amplitude is written by using these
five kinematical variables and three helicity-based form factors: $H_0(\qsq)$,
$H_+(\qsq)$, and $H_-(\qsq)$, which can be
computed by the lattice QCD. The helicity form factors are combinations of
one vector and two axial-vector form factors, which are parameterized
in general as,
\begin{equation}
A_i(\qsq) = \frac{A_i(0)}{1-\qsq/M_A^2},\ \ 
V_i(\qsq) = \frac{V_i(0)}{1-\qsq/M_V^2}.
\end{equation} 
Traditionally, three observables are used to describe the vector channel:
The branching fraction and the form factor ratios $\rv$ and $\rtwo$, which are
defined as $V(0)/A_1(0)$ and $A_2(0)/A_1(0)$, respectively.

\subsection{The s-wave interference in the $\dkpimn$ decay}
In 2002, the FOCUS experiment published a series of paper based on the $\dkpimn$
decay. A sample of 31,000 $\dkpimn$ events was reconstructed, providing one of
the largest $K^{*0}$ samples in the world. They updated the branching ratio and
the form factor measurements with an excellent precision, and
they discovered a surprising s-wave component in the decay, which was never
seen before in the charm semileptonic decays\cite{focusswave}.

When the distribution of the decay angle of $K^-$ in the $K^{-}\pi^+$ system
was analyzed, a huge forward-backward asymmetry was found, with its 
amplitude depending on the invariant mass of the $K^{-}\pi^+$. One possible
explanation was a quantum interference between the $K^{*0}$ and a s-wave 
component. They assumed a simple toy s-wave model
with a constant amplitude and a phase, and fitted the asymmetry. The
amplitude was measured about $7\%$ of the $K^{*0}$ Breit-Wigner amplitude
and the relative phase between the s-wave and the $K^{*0}$ was measured
at 45 degrees. Remarkably, the relative phase of the new s-wave
component were comparable to the one measured from a t-channel $K-\pi$
scattering experiment by the LASS Collaboration\cite{lass}. This
compatibility is not unexpected, since the semileptonic decays
do not involve final state interactions.  
  
\subsection{The measurement of the form factors in the $D^+$ decay}
The FOCUS experiment measured the form factors of the $\dksmn$ decays,
including the effects of the s-wave. The $\rv$ and $\rtwo$ were
obtained as $1.504 \pm 0.057 \pm 0.039$ and $0.875 \pm 0.049 \pm 0.064$,
respectively\cite{focusdff}. The FOCUS $\rv$ value is $2.9 \sigma$ away from 
the one measured by the E791 Collaboration, the previous world best\cite{e791}.
The updated world averages of the form factor ratios are $1.62 \pm 0.08$ and $0.83 \pm 0.05$, respectively.

\subsection{The measurement of the form factors in the $D_s$ decay}
According to the theoretical calculations, the form factors of the
$\dsphiln$ decay should be comparable to those of the $\dksln$ decay
within $10\%$. In the past, the $\rv$ measurement was consistent
between the $D_s$ and the $D^+$ decays, but the $\rtwo$ of the $D_s$ decay
was found twice the size of that of the $D^+$ decay.
Recently, the FOCUS experiment published a paper on the $\dsphimn$ 
decay\cite{focusdsphi}, where they measured the form factor ratios $\rv$ and
$\rtwo$ as $1.549 \pm 0.250 \pm 0.145$ and $0.713 \pm 0.202 \pm 0.266$,
respectively. Both values are consistent with those for the $\dksln$ decays.  

\section{The direct measurement of the $\Gamma(K^*\mu\nu)/\Gamma(K\mu\nu)$
ratio}
\begin{figure}\label{wills}
\begin{center}
\psfig{figure=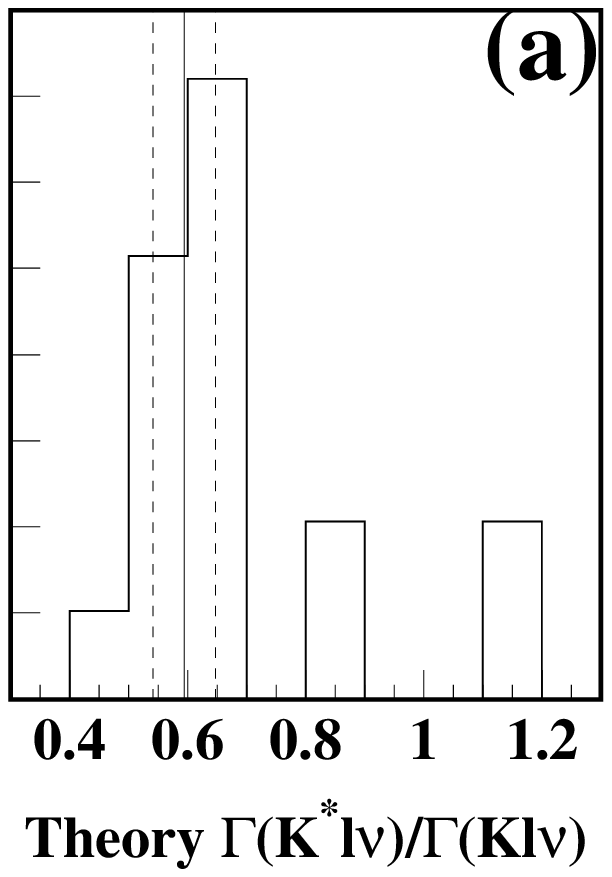,height=2.0in}
\hspace*{0.5in}
\psfig{figure=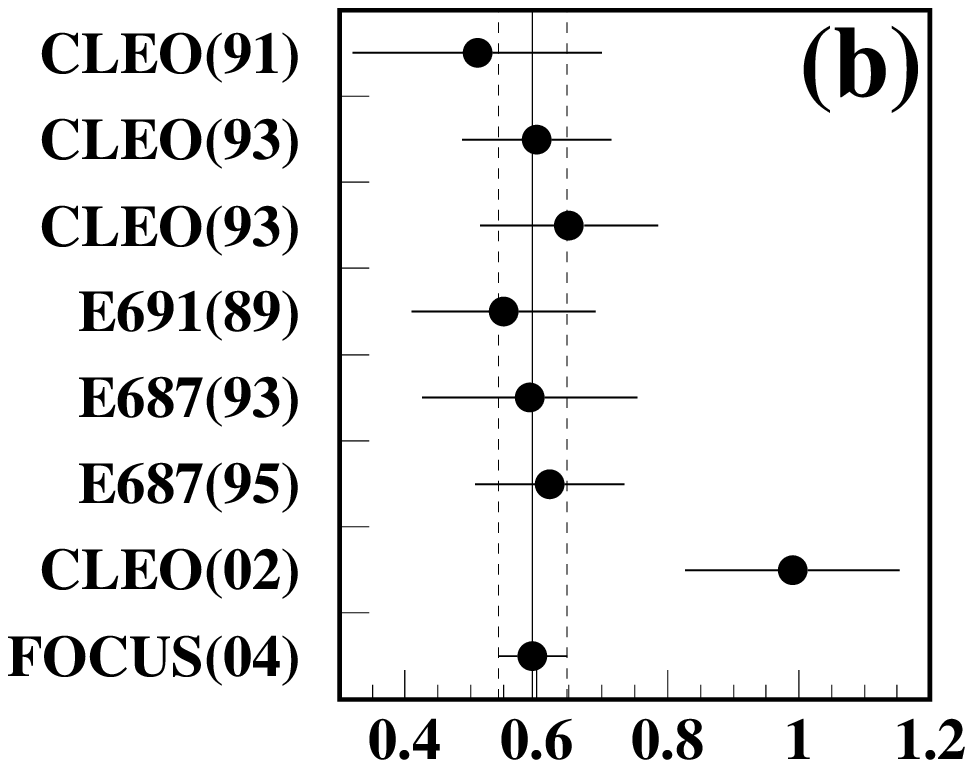,height=2.0in}
\end{center}
\caption{(a) The compilation of the theoretical calculations on the
relative branching ratio of $\Gamma(K^*l\nu)/\Gamma(Kl\nu)$.
(b) The compilation of the experimental measurements, including
both electron and muon channels. For both plots,
the sold lines and the dashed lines represent the value and the error
of the recent FOCUS measurement. The CLEO 2002 number was obtained
by using the PDG 2002 $\Gamma(\dkzen)$ value.
Courtesy of Prof.~Will Johns of Vanderbilt University.}
\end{figure}

In a naive model, the branching ratio of the vector charm semileptonic decays
(via a $K^*$) over the pseudoscalar ones (via a $K$) is about one. 
During the 90's, both the theoretical calculations and the experimental
measurements obtained smaller vector decay rate (See Figure~\ref{wills}.), but  
a year 2002 analysis by the CLEO measured the ratio near unity, which fact
generated an urgency to require further investigation. Recently, the FOCUS group
measured the ratio directly using the $\dksmn$ and $\dkzmn$ decays and
obtained $0.594 \pm 0.043 \pm 0.033$, confirming the measurements
from 90's\cite{focuswill}. The CLEO 2002 measurement was obtained
indirectly, by dividing the $\Gamma(\dksen)/\Gamma(\dkpipi)$ branching ratio
with the average $\Gamma(\dkzen)$ from the PDG 2000. The discrepancy between
the CLEO 2002 measurement and the other experimental measurements is
partly due to the slightly higher value for 
the $\Gamma(\dksen)/\Gamma(\dkpipi)$ and
partly due to the PDG $\Gamma(\dkzen)$ number, which is much lower
than the new FOCUS measurement, 
$\Gamma(\dkzmn) = 9.27 \pm 0.69 \pm 0.59 \pm 0.62$. 
 
\section{Summary}
Various results from the recent analysis on the charm semileptonic decays
were reviewed. In the pseudoscalar channel, the pole mass and the branching
ratio measurements are being updated by several experiments. 
In the vector channel, a new s-wave interference phenomena was found in the
$D^+$ decay, and the form factors were updated for both $D^+$ and $D_s$
decays, found to be consistent between two decay modes. The vector to
pseudoscalar decay ratio was updated with a consistent value to the 90's
measurements.

The CLEO-c experiment started data taking, showing a promising future and
other experiments are collecting high quality charm data sets.
We expect to resolve the various problems in the semileptonic charm physics
with excellent statistics in near future. 
 
\section*{Acknowledgments}
This work was supported by University of Illinois under contract
DE-FG02-91ER-40677 with the U.S. Department of Energy.

\end{document}